\documentclass{LMCS}

\def\dOi{11(3:13)2015}
\lmcsheading%
{\dOi}
{1--22}
{}
{}
{Nov.~14, 2014}
{Sep.~17, 2015}
{}

\ACMCCS{[{\bf Theory of computation}]: Formal languages and automata
  theory;  Theory and algorithms for application domains---Machine learning theory}

\usepackage{float} 
\usepackage{hyperref}

\usepackage{color,xcolor,graphicx}
\usepackage{caption, subcaption}
\usepackage{algorithm}
\usepackage{algpseudocode}

\usepackage{tikz,array}
\usetikzlibrary{arrows,automata}

\usepackage{bm}

\def    \S      {\Sigma}
\def    \d      {\delta}
\def    \e      {\epsilon}

\def    \se     {\subseteq}

\def    \N      {\mathbb{N}}
\def    \R      {\mathbb{R}}

\def    \B      {\mathbb{B}}
\def    \A      {\mathcal{A}}

\def    \ni     {\noindent}
\def    \ft#1   {\footnote{#1} }

\def    \comment#1 {}

\newcommand{\sy}[1]{{\bm{#1}}}  
\def    \a      {\sy{a}}
\def    \anew   {\a_{\text{\tiny new}}} 

\newcommand{\vx}    { \mathbf{x}}
\newcommand{\vy}    { \mathbf{y}}
\newcommand{\vz}    { \mathbf{z}}
\newcommand{\vzero} { \mathbf{0}}
\colorlet{colorb0}{black!10}
\colorlet{colorb1}{black!20}
\colorlet{colorb2}{black!10}
\colorlet{colorb3}{black!20}
\colorlet{colorb4}{black!10}
\colorlet{colorb5}{black!20}
\colorlet{colorb6}{black!40}
\colorlet{colorb7}{black!30}
\colorlet{colorb8}{black!20}

\colorlet{colorhighlight}{black!10}

\colorlet{colorconeA}{black!20}
\colorlet{colorconeB}{black!10}

\begin{document}

\title[Learning Regular Languages over Large Ordered Alphabets]
{Learning Regular Languages \\ over Large Ordered Alphabets}

\author[I-E. Mens]{Irini-Eleftheria Mens}
\address{VERIMAG \\  CNRS and University of Grenoble \\ France}
\email{\{irini-eleftheria.mens,oded.maler\}@imag.fr}

\author[O. Maler]{Oded Maler}


\titlecomment{This paper is an extended version of \cite{MalerMens14}}

\begin{abstract}
%
This work is concerned with regular languages defined over large alphabets, either infinite or just too large to be expressed enumeratively. We define a generic model where transitions are labeled by elements of a finite partition of the alphabet. We then extend Angluin's $L^*$ algorithm for learning regular languages from examples for such automata. We have implemented this algorithm and we demonstrate its behavior where the alphabet is a subset of the natural or real numbers. We sketch the extension of the algorithm to a class of languages over partially ordered alphabets. 
\end{abstract}

\maketitle

\section*{Introduction}\label{S:Intro}

The main contribution of this paper is a generic algorithm for learning regular languages defined over a large alphabet $\S$. Such an alphabet can be infinite, like $\N$ or $\R$ or just so large, like $\B^n$ for very large $n$ or large subsets of $\N$, so that it is impossible or impractical to treat it in an enumerative way, that is, to write down the entries of the transition function $\d(q,a)$ for every $a\in \S$. The obvious solution is to use a \emph{symbolic} representation where transitions are labeled by predicates which are applicable to the alphabet in question. Learning algorithms infer an automaton from a finite set of words (the \emph{sample}) for which membership is known. Over small alphabets, the sample should include the set $S$ of all the shortest words that lead to each state (access sequences) and, in addition, the set $S\cdot \S$ of all their $\S$-continuations. Over large alphabets this is not a practical option and as an alternative we develop a symbolic learning algorithm over \emph{symbolic words} which are only partially backed up by the sample. In a sense, our algorithm is a combination of automaton learning and learning of non-temporal predicates. Before getting technical, let us discuss briefly some motivation.

Finite automata are among the corner stones of Computer Science. From a practical point of view they are used routinely in various domains ranging from syntactic analysis, design of user interfaces or administrative procedures to implementation of digital hardware and verification of software and hardware protocols. Regular languages admit a very nice, clean and comprehensive theory where different formalisms such as automata, logic, regular expressions, semigroups and grammars are shown to be equivalent. The problem of learning automata from examples was introduced already in 1956 by Moore \cite{moore1956gedanken}. This problem, like the problem of  automaton minimization,  is closely related to the Nerode right-congruence relation over words associated with every language or sequential function \cite{Nerode58}. This relation declares two \emph{input histories} as equivalent if they lead to the same \emph{future continuations}, thus providing  a crisp characterization of what a \emph{state} in a dynamical system is in terms of observable input-output behavior. All algorithms for learning automata from examples, starting with the seminal work of Gold \cite{Gold72} and culminating in the well-
known $L^*$ algorithm of Angluin \cite{Angluin87} are based on this concept \cite{dehiguera2010book}.

One weakness, however, of the classical theory of regular languages is that it is rather ``thin" and ``flat". In other words, the alphabet is often considered as a small set devoid of any additional structure. On such alphabets, classical automata are good for expressing and exploring the temporal (sequential, monoidal) dimension embodied by the concatenation operations, but  less good in expressing ``horizontal" relationships. To make this statement  more concrete, consider the verification of a system consisting of  $n$ automata running in parallel, making independent as well as synchronized transitions. To express the set of joint behaviors of this product of automata as a formal language, classical theory will force you to use the exponential alphabet of global states and indeed, a large part of verification is concerned with fighting this explosion using constructs such as BDDs and other logical forms that exploit the sparse interaction among components. This is done, however, without a real interaction with classical formal language theory (one exception is the theory of \emph{traces} \cite{Traces} which attempts to treat this issue but in a very restricted context).

These and other considerations led us to use \emph{symbolic automata} as a generic framework for recognizing languages over large alphabets where transitions outgoing from a state are labeled, semantically speaking, by \emph{subsets} of the alphabet.  These subsets are expressed syntactically according to the specific alphabet used:  Boolean formulae when $\S=\B^n$ or by some classes of inequalities when $\S\subseteq\N$ or $\S\subseteq \R$. Determinism and completeness of the transition relation, which are crucial for learning and minimization, can be enforced by requiring that the subsets of $\S$ that label the transitions outgoing from a given state form a \emph{partition} of the alphabet. Such symbolic automata have been used in the past for Boolean vectors \cite{mona95} and have been studied extensively in recent years as acceptors and  transducers where transitions are guarded by predicates of various theories  \cite{Veanes-vmcai11,Veanes-symb}.

Readers working on program verification or hybrid automata are, of course, aware of automata with symbolic transition guards but it should be noted that in the model that we use, \emph{no auxiliary variables} are added to the automaton. Let us stress this point by looking at a popular extension of automata to infinite alphabets, initiated in  \cite{KaminskiF94} using \emph{register automata} to accept \emph{data languages} (see \cite{benedikt2010} for a good exposition of theoretical properties and \cite{HowarSJC12} for learning algorithms). In that framework, the automaton is augmented with additional registers that can store some input letters. The registers can then be compared with newly-read letters and influence transitions.  With register automata one can express, for example, the requirement that the password at login is the same as the password at sign-up. This very restricted use of memory makes register automata much simpler than more notorious automata with variables whose emptiness problem is typically undecidable. The downside is that beyond \emph{equality} they do not really exploit the potential richness of the alphabets and their corresponding theories.

Our approach is different:  we do allow the \emph{values} of the input symbols to influence transitions via predicates, possibly of a restricted complexity. These predicates involve domain \emph{constants} and they partition the alphabet into finitely many classes. For example, over the integers a state may have transitions labeled by conditions of the form $c_1\leq x \leq c_2$  which give real (but of limited resolution) access to the input domain. On the other hand, we insist on a finite (and small) memory so that the exact value of $x$ \emph{cannot} be registered and has no future influence beyond the transition it has triggered.  Many control systems, artificial (sequential machines working on quantized numerical inputs) as well as natural (central nervous system, the cell),  are believed to operate in this manner. The automata that we use, like the  symbolic automata and transducers studied in \cite{Veanes-vmcai11,Veanes-symb,Veanes-SAtoolkit}, are geared toward languages recognized by automata having a large alphabet and a relatively-small state space.

We then develop a symbolic version of Angluin's $L^*$ algorithm for learning regular sets from queries and counter-examples whose output is a symbolic automaton. The main difference relative to the concrete algorithm is that in the latter, every transition $\d(q,a)$  in a conjectured automaton has at least one word in the sample that exercises it. In the symbolic case, a transition $\d(q,\sy{a})$ where $\sy{a}$ stands for a \emph{set} of concrete symbols, will be backed up in the sample only by a \emph{subset} of $\sy{a}$. Thus, unlike concrete algorithms where a counter-example always leads to a discovery of one or more new states, in our algorithm it may sometimes only modify the boundaries between partition blocks without creating new states. There are some similarities between our work and another recent adaptation of the $L^*$ algorithm to symbolic automata, the $\S^*$ algorithm of \cite{bb13sigma}. This work is incomparable to ours as they use a richer model of transducers and more general predicates on inputs and outputs. Consequently their termination result is weaker and is relative to the termination of the counter-example guided abstraction refinement procedure.

The rest of the paper is organized as follows. In Section~\ref{S:GeneralLearning} we provide a quick summary of learning algorithms over small alphabets. In Section~\ref{S:SymbolicAutomata} we define symbolic automata and then extend the structure which underlies all automaton learning algorithms, namely the \emph{observation table}, to be symbolic, where symbolic letters represent sets, and where entries in the table are supported only by partial evidence. In Section~\ref{S:SymboliLearningTotalOrder} we write down a symbolic learning algorithm, an adaptation of $L^*$ for totally ordered alphabets such as $\R$ or $\N$ and illustrate the behavior of a prototype implementation. The algorithm is then extended to languages over partially ordered alphabets such as $\N^d$ and $\R^d$ where in each state, the labels of outgoing transition from a monotone partition of the alphabet are represented by finitely many points.  We conclude by a discussion of past and future work.

\section{Learning Regular Sets}\label{S:GeneralLearning}

We briefly survey Angluin's $L^*$ algorithm \cite{Angluin87} for learning regular sets from membership queries and counter-examples, with slightly modified definitions to accommodate for its symbolic extension. Let $\S$ be a finite alphabet and let $\S^*$ be the set of sequences (words) over $\S$. Any order relation $<$ over $\S$ can be naturally lifted to a lexicographic order over $\S^*$. \comment{ that we denote by $<_l$.} With a language $L\se \S^*$ we associate a \emph{characteristic function} $f:\S^*\to\{+,-\}$, where $f(w) = +$ if the word $w \in \S^{*}$ belongs to $L$ and $f(w) = -$, otherwise. 

A \emph{deterministic finite automaton}  over $\S$ is a tuple $\A = (\S, Q, \d,
q_0,F)$, where $Q$ is a non-empty finite set of \emph{states}, $q_0 \in Q$ is
the \emph{initial} state, $\d : Q \times \S \to Q$ is the \emph{transition
function}, and $F \se Q$ is the set of \emph{final} or \emph{accepting} states.
The transition function $\d$ can be extended to $\d: Q \times \S^{*} \to Q$,
where $\d(q,\e) = q$, and $\d(q,u \cdot a) = \d(\d(q,u),a)$ for $q \in Q$, $a
\in \S$ and $u \in \S^{*}$. A word $w\in \S^{*}$ is \emph{accepted} by $\A$ if
$\d(q_0,w) \in F$, otherwise $w$ is \emph{rejected}. The language recognized by
$\A$ is the set of all accepted words and is denoted by $L(\A)$. 

Learning algorithms, represented by the \emph{learner}, are designed to infer
an unknown regular language $L$ (the \emph{target language}). The learner aims
to construct a finite automaton that recognizes $L$ by gathering
information from the \emph{teacher}. The \emph{teacher} knows $L$ and can
provide information about it. It can answer two types of queries:
\emph{membership queries}, i.e., whether a given word belongs to the target
language, and \emph{equivalence queries}, i.e., whether a conjectured automaton
suggested by the learner is the right one. If this automaton fails to accept $L$
the teacher responds to the equivalence query by a \emph{counter-example}, a
word miss-classified by the conjectured automaton.

In the $L^*$ algorithm, the learner starts by asking membership queries. All information provided is suitably gathered in a table structure, the \emph{observation table}. Then, when the information is sufficient, the learner constructs a \emph{hypothesis automaton} and poses an equivalence query to the teacher. If the answer is positive then the algorithm terminates and returns the conjectured automaton. Otherwise the learner accommodates the information provided by the counter-example into the table, asks additional membership queries until it can suggest a new hypothesis and so on, until termination.

A prefix-closed set $ S \uplus R \subset \S^{*} $ is a \emph{balanced $\S$-tree} if $\forall a \in \S$: 1) For every $s \in S$ $s\cdot a \in S\cup R$, and 2)  For every $r\in R$,  $r \cdot a \not \in S\cup R $. Elements of $R$ are called \emph{boundary elements} or \emph{leaves}.
\footnote{We use $\uplus$ for disjoint union.}

\begin{defi} [Observation Table]
An \emph{observation table} is a tuple $T = ( \S, S, R, E, f )$ such that $\S$
is an alphabet, $S \cup R$ is a balanced $\S$-tree, $E$ is a subset of
$\S^{*}$ and $f : (S \cup  R) \cdot E \to \{-, +\}$ is the classification
function, a restriction of the characteristic function of the target language
$L$.
\end{defi}

\ni The set $(S \cup  R) \cdot E$ is the \emph{sample} associated with the
table, that is, the set of words whose membership is known. The elements of $S$
admit a tree structure isomorphic to a \emph{spanning tree} of the transition
graph rooted in the initial state. Each $s\in S$ corresponds to a state $q$ of
the automaton for which $s$ is an \emph{access sequence}, one of the shortest
words that lead from the initial state to  $q$. The elements of $R$ should tell
us about the back- and cross-edges in the automaton and the elements of $E$ are 
``experiments" that should be sufficient to distinguish between states. This
works by associating with every  $s \in S \cup R$ a specialized classification
function $ f_s : E \to \{ - , + \} $, defined as $ f_s (e) = f( s \cdot e)$,
which characterizes the row of the observation table labeled by $s$. To build an
automaton from a table it should satisfy certain conditions.

\begin{defi}[Closed, Reduced and Consistent Tables]
An observation table $T$ is:
    \begin{itemize}
        \item Closed if for every $r \in R$, there exists an $s \in S$, such that $f_r = f_s$;
        \item Reduced if for every $s,s'\in S~f_s\not=f_{s'}$;
        \item Consistent if for every $s, s' \in S$,  $f_{s} = f_{s'}$ implies 
        $ f_{s\cdot a }  = f_{ s'\cdot a }, \forall a \in \S$.
    \end{itemize}
\end{defi}
\ni Note that a reduced table is trivially consistent and that for a closed and
reduced table we can define a function $g:R\to S$ mapping every $r\in R$ to the
unique $s\in S$ such that $f_s=f_r$. From such an observation table $T= ( \S, S,
R, E, f )$ one can construct an automaton $\A_T = (\S, Q, q_0, \d, F)$ where $Q
=S$, $q_0=\e$, $F=\{s \in S:f_s(\e) = +\}$ and $$ \d(s,a) = \left\{
\begin{array}{l l} s \cdot a   &  \mbox{ when } s \cdot a \in S\\    g(s\cdot a)
  & \mbox{ when } s \cdot a  \in R \end{array} \right.$$

\ni The learner attempts to keep the table closed at all times.
The table is not closed when there is some $r \in R $ such that $ f_{r}$ is different from $f_{s}$ for all $s \in S $. To close the table, the learner moves $r$ from $R$ to $S$ and adds the $\S$-successors of $r$, i.e., all words $r\cdot a$ for $a\in\S$, to $R$. The extended table is then filled up by asking membership queries until it becomes closed.


Variants of the $L^*$ algorithm differ in the way they treat counter-examples, as described in more detail in \cite{BergR04}. The original algorithm \cite{Angluin87} adds all the \emph{prefixes} of the counter-example to $S$ and thus possibly creating inconsistency that should be fixed. The version proposed in \cite{maler1995learnability} for learning $\omega$-regular languages adds all the \emph{suffixes} of the counter-example to $E$. The advantage of this approach is that the table always remains consistent and reduced with $S$ corresponding exactly to the set of states. A disadvantage is the possible introduction of redundant columns that do not contribute to further discrimination between states. The symbolic algorithm that we develop in this paper is based on an intermediate variant, referred to in \cite{BergR04} as the \emph{reduced observation algorithm}, where some prefixes of the counter-example are added to $S$ and some suffixes are added to $E$.


\begin{exa} \label{ex:Lstar}
We illustrate the behavior of the $L^{*}$ algorithm while learning a language
$L$ over $\S = \{1,2,3,4,5\}$. We use the tuple $(w,+)$ to indicate a counter-example $w \in
L$ rejected by the conjectured automaton, and $(w,-)$ for the opposite case.
Initially, the observation table is $T_0 = (\S, S, R, E, f)$ with $S = E =
\{\epsilon\}$ and $R = \S$ and we ask membership queries for all words in $(S
\cup R) \cdot E$ to obtain table $T_0$, shown in Fig.~\ref{fig:Angluintables}.
The table is not closed so we move word $1$ to $S$, add its continuations,
$1\cdot\S$ to $R$ and ask membership queries to obtain table $T_1$ which is
now closed. We construct an hypothesis $\A_1$ (Fig.~\ref{fig:AngluinsHA}) from
this table, and pose an equivalence query for which the teacher returns 
counter-example $(3\cdot1,-)$. We add $3\cdot1$ and its prefix $3$ to set $S$ and
add all their continuations to the boundary of the table resulting table
$T_2$ of Fig.~\ref{fig:Angluintables}. This table is not consistent:
two elements $\epsilon$ and $3$ in $S$ are equivalent but their successors 
$1$ and $3\cdot1$ are not. In order to distinguish the two strings we add to $E$
the suffix $1$ and end up with a closed and consistent table $T_3$. The new
hypothesis for this table is $\A_3$, shown in Fig.~\ref{fig:AngluinsHA}. Once 
more the equivalence query will return a counter-example, $(1\cdot3\cdot3,-)$. We
again add the counter-example and prefixes to the table,  ask membership queries
to fill in the table and solve the inconsistency that appears for $1$ and
$1\cdot3$ by adding suffix 3 to the table. The table corresponds now to the
correct hypothesis $\A_5$, and the algorithm
terminates. \qed
\end{exa}

\begin{figure}
    \centering
    \scalebox{.8}{

\begin{tabular*}{.98\textwidth}{|@{ \extracolsep{\fill} }c c c c c c@{ }|}
    \hline
    $T_0$ & $T_1$ & $T_2$ & $T_3$ & $T_4$ & $T_5$ \\
    \begin{tabular}{|@{ }r@{ }|@{ }c@{ }|}
        \hline
         & $\epsilon$ \\
        \hline
        $\epsilon$ & - \\
        \hline
        $1$ & + \\
        $2$ & + \\
        $3$ & - \\
        $4$ & - \\
        $5$ & - \\
        \hline
    \end{tabular}
    &
    \begin{tabular}{|@{ }r@{ }|@{ }c@{ }|}
        \hline
         & $\epsilon$ \\
        \hline
        $\epsilon$ & - \\
        $1$ & + \\
	\hline 
        $2$ & + \\
        $3$ & - \\
        $4$ & - \\
        $5$ & - \\
        $1\cdot1$ & - \\
        $1\cdot2$ & - \\
        $1\cdot3$ & + \\
        $1\cdot4$ & - \\
        $1\cdot5$ & - \\
       \hline
    \end{tabular}
    &
    \begin{tabular}{|@{ }r@{ }|@{ }c@{ }|}
        \hline
         & $\epsilon$ \\
        \hline
        $\epsilon$ & - \\
        $1$  & + \\
        $3$  & - \\
        $3\cdot1$ & - \\
	\hline 
        $2$  & + \\
        $4$  & - \\
        $5$  & - \\
        $1\cdot1$ & - \\
        $1\cdot2$ & - \\
        $1\cdot3$ & + \\
        $1\cdot4$ & - \\
        \vdots   \\ 
    \end{tabular}  
    &
    \begin{tabular}{|@{ }r@{ }|@{ }c c @{ }|}
        \hline
         & $\epsilon$ & $1$ \\
        \hline
        $\epsilon$ & - & + \\
        $1$  & + & - \\
        $3$  & - & - \\
        $3\cdot1$ & - & - \\
	\hline 
        $2$  & + & - \\
        $4$  & - & - \\
        $5$  & - & - \\
        $1\cdot1$ & - & - \\
        $1\cdot2$ & - & - \\
        $1\cdot3$ & + & - \\
        $1\cdot4$ & - & - \\ 
        \vdots   \\ 
    \end{tabular}   
    &
    \begin{tabular}{|@{ }r@{ }|@{ }c c @{ }|}
        \hline
         & $\epsilon$ & $1$ \\
        \hline
        $\epsilon$ & - & + \\
        $1$  & + & - \\
        $3$  & - & - \\
        $1\cdot3$ & + & - \\
        $3\cdot1$ & - & + \\
        $1\cdot3\cdot3$ & - & - \\
	\hline 
        $2$  & + & - \\
        $4$  & - & - \\
        $5$  & - & - \\
        $1\cdot1$ & - & - \\
        $1\cdot2$ & - & - \\
        \vdots   \\ 
    \end{tabular}   
    &
    \begin{tabular}{|@{ }r@{ }|@{ }c c c@{ }|}
        \hline
         & $\epsilon$ & $1$ & $3$\\
        \hline
        $\epsilon$ & - & + & - \\
        $1$  & + & - & + \\
        $3$  & - & - & - \\
        $1\cdot3$ & + & - & - \\
        $3\cdot1$ & - & + & - \\
        $1\cdot3\cdot3$ & - & - & - \\
	\hline 
        $2$  & + & - & + \\
        $4$  & - & - & - \\
        $5$  & - & - & - \\
        $1\cdot1$ & - & - & - \\
        $1\cdot2$ & - & - & - \\
        \vdots   \\ 
    \end{tabular} \\
& & & & & \\
    \hline
\end{tabular*}}
    \smallskip
    \caption{Observation tables for Example~\ref{ex:Lstar}.}
    \label{fig:Angluintables}
\end{figure}

\begin{figure}
    \centering
    \scalebox{.9}{
    \begin{tabular}{|@{ \extracolsep{\fill} } c | c | c @{ }|}
        \hline
        $\A_1$ & $\A_3$ & $\A_5$\\
	\begin{tabular}{c}
        \input{./Figures/Ex_Lstar/ex_Lstar_h0}
	\end{tabular}
	    &   
	\begin{tabular}{c}
	\input{./Figures/Ex_Lstar/ex_Lstar_h1} 
	\end{tabular}
	    &   
	\begin{tabular}{c}
	\input{./Figures/Ex_Lstar/ex_Lstar_h2} 
	\end{tabular}\\
        \hline\noalign{\smallskip}
    \end{tabular}}
    \caption{Hypotheses for Example~\ref{ex:Lstar}}
    \label{fig:AngluinsHA}
\end{figure}



\section{Symbolic Automata}\label{S:SymbolicAutomata}

In this section we introduce the variant of \emph{symbolic automata} that we use. Symbolic automata \cite{Veanes-vmcai11,Veanes-SAtoolkit} give a more
succinct representation for languages over large finite alphabets and can also
represent languages over infinite alphabets such as $\N$, $\R$, or
$\R^{n}$.
The size of a standard automaton for a language grows linearly with the size of the
alphabet and so does the complexity of learning algorithms such as $L^{*}$.
As we shall see, symbolic automata admit a variant of the $L^{*}$ algorithm
whose complexity is independent of the alphabet size.

Let $\S$ be a large, possibly infinite, alphabet, to which we will refer from now on as the
\emph{concrete} alphabet. We define a symbolic automaton to be an
automaton over $\S$ where each state has a small number of outgoing
transitions labeled by symbols that represent subsets of $\S$. For every state, these subsets form a (possibly different) \emph{partition} of $\S$ and hence the automaton is complete and deterministic. We start with an arbitrary alphabet viewed as an unstructured set and
present the concept in purely semantic manner before we move to ordered sets
and inequalities in subsequent sections.

Let $\sy{\S}$ be a finite alphabet, that we call the \emph{symbolic alphabet} and its elements \emph{symbolic letters} or \emph{symbols}.
Let $\psi:\S \to \sy{\S}$ map concrete letters into symbolic ones. The $\S$-semantics of a \emph{symbolic letter} $\a\in \sy{\S}$ is defined as
$[\a]_\psi=\{a \in \S: \psi(a)=\a\}$ and the set $\{[\a]_\psi: \a \in \sy{\S}\}$ forms a \emph{partition} of $\S$. We will often omit $\psi$ from the notation and use $[\a]$ when $\psi$, which is always present, is clear from the context. The $\S$-semantics can be extended to symbolic words of the form $\sy{w} =
\a_1 \cdot \a_2 \cdots \a_k\in \sy{\S}^*$ as the concatenation of the concrete one-letter languages associated with the respective symbolic letters or, recursively speaking, $[\sy{\e}]=\{\e\}$ and $[\sy{w}\cdot \a]=[\sy{w}]\cdot [\a]$ for $\sy{w}\in \sy{\S}^*$, $\a\in\sy{\S}$.

\begin{defi}[Symbolic Automaton]
    \label{def:symb-aut}
    A \emph{deterministic symbolic automaton}  is a tuple
    $\sy{\A} = (\S,\sy{\S},\psi, Q, \d, \sy{\d}, q_0, F)$, where
     \begin{itemize}
         \item $\S$ is the input alphabet,
         \item $\sy{\S}$ is a finite alphabet, decomposable into
                $\sy{\S}=\biguplus _{q\in Q}\sy{\S}_q$,
         \item $\psi=\{\psi_q:q\in Q\}$ is a family of surjective functions
                $\psi_q: \S \to \sy{\S}_q$,
         \item $Q$ is a finite set of states,
         \item $\d : Q \times \S \to Q$ and $\sy{\d}:Q \times \sy{\S} \to Q$
		are the concrete and symbolic transition functions respectively,
		such that $\d(q,a) = \sy{\d}(q,\psi_q(a))$,
         \item $q_0$ is the initial state and $F$ is a set of accepting states.
    \end{itemize}
\end{defi}

\ni The transition function is extended to words as in the concrete case and the
symbolic automaton can be viewed as an acceptor of a concrete language. When at
$q$ and reading a concrete letter $a$, the automaton will take the transition
$\sy{\d}(q,\a)$ where $\a$ is the \emph{unique} element of $\sy{\S}_q$
satisfying $a\in [\a]$. Hence $L(\sy{\A})$ consists of all concrete words
whose run leads from $q_0$ to a state in $F$. A language $L$ over alphabet $\S$
is symbolic recognizable if there exists a symbolic automaton $\sy{\A}$ such
that $L = L(\sy{\A})$.

\ni {\bf Remark}: The association of a \emph{symbolic language} with a symbolic automaton is more subtle because we allow different partitions of $\S$ and hence different symbolic input alphabets at different states. The transition to be taken while being in a state $q$ and reading a symbol $\a\not \in \sy{\S}_q$ is well defined only when $[\a]\se [\sy{a'}]$ for some $\sy{a'}\in \sy{\S}_q$.  Such a model can be transformed into an automaton which is complete over a  symbolic alphabet which is common to all states as follows.  Let $$\sy{\S}'=\prod_{q\in Q} \sy{\S}_q, \mbox{~with the $\S$-semantics~~}  [(\a_1,\ldots,\a_n)]=[\sy{a_1}]\cap \ldots \cap [\a_n],$$ and let $\tilde{\sy{\S}}=\{\sy{b}\in \sy{\S}':[\sy{b}]\not = \emptyset\}$. Then we define $\sy{\widetilde{\A}}=(\tilde{\sy{\S}}, Q, \tilde{\d}, q_0, F)$ where, by construction, for every $\sy{b}\in \tilde{\sy{\S}}$ and every $q\in Q$, there is a unique $\a\in \sy{\S}_q$ such that $[\sy{b}]\se [\a]$ and hence one can define
the transition function as $\tilde{\sy{\d}}(q,\sy{b})=\sy{\d}(q,\a)$. This model is more comfortable for language-theoretic studies but in the learning context it introduces an unnecessary blow-up in the alphabet size and the number of queries for every state. For this reason we stick in this
paper to the Definition~\ref{def:symb-aut} which is more economical. A similar approach of state-local abstraction has been taken in \cite{local-alph} for learning parameterized language. The construction of $\sy{\S}'$ is similar to the minterm construction of \cite{Veanes-minimization} used to create a common alphabet in order to apply the minimization algorithm of Hopcroft to symbolic automata. Anyway, in our learning framework symbolic automata are used to read concrete and not symbolic words. \qed

It is straightforward that for a finite concrete alphabet $\S$ the set of languages accepted by symbolic automata coincides with the set of recognizable
regular languages over $\S$. Moreover, even when the alphabet is infinite, closure under Boolean operations is preserved.

\begin{prop}[Closure under Boolean Operations]
  Languages accepted by deterministic symbolic automata are effectively closed
under Boolean operations.
\end{prop}

\proof
Closure under complement is immediate by complementing the set of accepting
states. For intersection the standard product construction is adapted as follows.
Let $L_1,L_2$ be languages recognized by the symbolic automata
$\sy{\A}_1=(\S,\sy{\S}_1,\psi_1,Q_1,\d_1, \sy{\d}_1,q_{01},F_1)$, and
$\sy{\A}_2=(\S,\sy{\S}_2,\psi_2,Q_2,\d_2, \sy{\d}_2,q_{02},F_2)$, respectively.
Let $\sy{\A}= (\S,\sy{\S},\psi,Q,\d, \sy{\d},q_{0},F)$, where
\begin{itemize}
    \item $Q=Q_1\times Q_2$, $q_0=(q_{01},q_{02})$, $F=F_1\times F_2$,
    \item For every $(q_1,q_2)\in Q$
    \begin{itemize}
        \item $\sy{\S}_{(q_1,q_2)} = \{(\a_1,\a_2) \in \sy{\S}_1
                \times \sy{\S}_{2} \mid [\a_1] \cap [\a_2] \neq \emptyset
\}$
        \item $\psi_{(q_1,q_2)}(a) = (\psi_{1,q_1}(a),\psi_{2,q_2}(a))$,
                $\forall a\in \S$
        \item $\sy{\d}((q_1,q_2),(\a_1,\a_2)) =
                (\sy{\d}_1(q_1,\a_1),\sy{\d}_2(q_2,\a_2))$,
                $\forall (\a_1,\a_2)\in \sy{\S}_{(q_1,q_2)}$
    \end{itemize}
\end{itemize}
It is sufficient to observe that the corresponding implied concrete automata
$\A_1$, $\A_2$ and $\A$ satisfy $\d((q_1,q_2),a)= (\d_1(q_1,a),\d_2(q_2,a))$ and
the standard proof that $L(\A)=L(\A_1)\cap L(\A_2)$ follows. Closure under union and set difference is then evident.  \qed
The above product construction is used to implement equivalence queries where both the target language and the current conjecture are represented by symbolic automata. A counter-example is found by looking for a shortest path in the product automaton from the initial state to a state in $F_1\times (Q_2-F_2) \cup (Q_1-F_1)\times F_2$ and selecting a lexicographically minimal concrete word along that path.

\begin{figure}[ht]
\begin{subfigure}[b]{.6\textwidth}
\centering
\scalebox{.85}{ \input{./Figures/Ex_SA/exSA} }
\end{subfigure}
\hfill
\begin{subfigure}[b]{.38\textwidth}
\centering
  \begin{tabular}{| r | c c c c |}
   \hline
   $\sy{\d}$ & $q_0$ & $q_1$ & $q_2$ & $q_3$ \\
   \hline
   $q_0$ & $-$        & $\a_0$ & $\a_1$ & $-$ \\
   $q_1$ & $-$        & $-$        & $\a_3$ & $\a_2$ \\
   $q_2$ & $\a_4$ & $\a_6$ & $\a_7$ & $\a_5$ \\
   $q_3$ & $-$        & $-$        & $\a_8$ & $-$ \\
   \hline
\end{tabular}
  \vspace{2.5em}
\end{subfigure}
 \caption{A symbolic automaton $\sy{\A}$ with its symbolic transition function.}
 \label{Fig:ExampleSymbolicAutomaton}
\end{figure}

\begin{exa}\label{ex:1dim}
Figure~\ref{Fig:ExampleSymbolicAutomaton} shows a symbolic automaton equivalent
to automaton $\A_5$ of Figure~\ref{fig:AngluinsHA}.
The symbolic alphabets for the states are
$\sy{\S}_{q_0} = \{\a_0, \a_1\}$,
$\sy{\S}_{q_1} = \{\a_2, \a_3\}$,
$\sy{\S}_{q_2} = \{\a_4, \a_5, \a_6, \a_7\}$,
$\sy{\S}_{q_3} = \{\a_8\}$,
and the $\S$-semantics for the symbols is
$[\a_0] = \{1,2\}$,
$[\a_1] = \{3,4,5\}$,
$[\a_2] = \{3\}$,
$[\a_3] = \{1,2,4,5\}$, etc..
The same automaton can accept a language over the uncountable alphabet
$\S = [0,100) \subset \R$, defining $\psi$ as shown in Figure~\ref{Fig:ExamplePsiFunction1dim}.
\end{exa}

\begin{figure}[ht]
 \centering
\scalebox{.85}{\input{./Figures/Ex_SA/exSA_psi}}
 \caption{The concrete semantics of the symbols of automaton $\sy{\A}$ of Fig.~\ref{Fig:ExampleSymbolicAutomaton}, when defined over $\S = [0,100) \subseteq \R$.}
 \label{Fig:ExamplePsiFunction1dim}
\end{figure}

\section{Symbolic Observation Tables}\label{S:SymboliLearning}

In this section we adapt observation tables to the symbolic setting. They are
similar to the concrete case with the additional notions  of evidences and
evidence compatibility.

\begin{defi}[Balanced Symbolic $\S$-Tree]
    A \emph{balanced symbolic $\S$-tree} is a  tuple
$(\sy{\S},\sy{S},\sy{R}$, $\psi)$ where
    \begin{itemize}
        \item $\sy{S}\uplus \sy{R}$ is a prefix-closed subset of $\sy{\S}^{*}$
        \item $\sy{\S}=\biguplus_{\sy{s}\in \sy{S}}\sy{\S}_{\sy{s}}$ is a
                symbolic alphabet
        \item $\psi=\{\psi_{\sy{s}}\}_{\sy{s} \in \sy{S}}$
                is a family of total surjective functions of the form
                $\psi_{\sy{s}}:\S\to\sy{\S_s}$.
    \end{itemize}
It is required that for every $\sy{s} \in \sy{S}$ and $\a\in
\sy{\S}_{\sy{s}}$, $\sy{s}\cdot \a\in \sy{S}\cup \sy{R}$  and for any
$\sy{r} \in \sy{R}$ and $\a\in\sy{\S}$, $\sy{r}\cdot \a\not
\in\sy{S}\cup \sy{R}$ . Elements of $\sy{R}$ are called boundary elements of the
tree.
\end{defi}

\ni We will use observation tables whose rows are symbolic words and hence an
entry in the table will constitute a statement about the inclusion or exclusion
of a large \emph{set} of concrete words in the language. We will not ask
membership queries concerning all those concrete words, but only for a small
representative subset that we call \emph{evidence}.

\begin{defi}[Symbolic Observation Table]
A \emph{symbolic observation table}  is a tuple
$\sy{T}=(\S, \sy{\S}, \sy{S}, \sy{R}, \psi, E, \sy{f}, \mu)$ such that
    \begin{itemize}
        \item $\S$ is an alphabet,
        \item $(\sy{\S},\sy{S},\sy{R},\psi)$  is a balanced symbolic
                $\S$-tree (with $\sy{R}$ being its \emph{boundary}),
        \item $E$ is a subset of $ \S^{*}$,
        \item  $\sy{f} : ( \sy{S} \cup \sy{R} ) \cdot E \to \{-,+\}$
                is the symbolic classification function
        \item  $\mu : (\sy{S} \cup \sy{R} ) \cdot E \to
                2^{\S^{*}}-\{\emptyset\}$
                is an evidence function satisfying
                $\mu(\sy{w}) \subseteq [\sy{w}]$.
                The image of the evidence function is prefix-closed:
                $w\cdot a \in \mu(\sy{w}\cdot \a) \Rightarrow w \in
                \mu(\sy{w})$.
    \end{itemize}
\end{defi}

\ni As for the concrete case we use $ \sy{f}_{\sy{s}} : E \to \{-,+\}$ to
denote the partial evaluation of $\sy{f}$ to some symbolic word $ \sy{s} \in
\sy{S} \cup \sy{R}$, such that, $ \sy{f}_{\sy{s}}(e) = \sy{f}(\sy{s}\cdot  e)$.
Note that the set $E$ consists of \emph{concrete} words but this poses no
problem because elements of $E$ are used only to distinguish between states and
do not participate in the derivation of the symbolic automaton from the table.
Concatenation of a symbolic word and a concrete one follows concatenation of
symbolic words as defined above where each concrete letter $a$ is considered as
a symbolic letter $\a$ with $[\a] = \{a\}$ and $\mu(\a) = a$.
The notions of closed, consistent and reduced table are similar to the concrete
case.

The set $\sy{M}_{\sy{T}} = (\sy{S} \cup \sy{R}) \cdot E$ is called the
\emph{symbolic sample} associated with $\sy{T}$. We require that for each word
$\sy{w}\in \sy{M}_{\sy{T}}$ there is at least one concrete $w\in \mu(\sy{w})$
whose membership in $L$, denoted by $f(w)$, is known. The set of such words is
called the \emph{concrete sample} and is defined as $M_{\sy{T}}=\{s\cdot e: s\in
\mu(\sy{s}), \sy{s}\in \sy{S}\cup\sy{R}, e\in E\}$. A table where  all evidences
of the same symbolic word admit the same classification is called
\emph{evidence-compatible}.

\begin{defi}[Table Conditions]
A table $\sy{T}=(\S, \sy{\S}, \sy{S}, \sy{R}, \psi, E, \sy{f}, \mu)$ is
    \begin{itemize}
        \item Closed if $ \forall \sy{r} \in \sy{R}$,
                $\exists \sy{s}=g(\sy{r}) \in \sy{S}$, $\sy{f}_{\sy{r}} =
                \sy{f}_{\sy{s}}$,
        \item Reduced if $\forall \sy{s}, \sy{s'} \in \sy{S}$, $\sy{f}_{\sy{s}}
                \neq \sy{f}_{\sy{s'}}$,
        \item Consistent if $\forall \sy{s}, \sy{s}' \in \sy{S}$,  $\sy{f_{s} =
                f_{s'}}$ implies
                $ \sy{f_{s\cdot a }  = f_{s'\cdot a }}, \forall \a \in
                \sy{\S_s}$.
        \item Evidence compatible if $\forall \sy{w} \in \sy{M}_{\sy{T}}$,
                $\forall w_1,w_2 \in \mu (\sy{w}), f( w_1 ) = f( w_2 )$.
    \end{itemize}
\end{defi}
\ni When a table $\sy{T}$ is evidence compatible the symbolic classification
function $\sy{f}$ can be defined for every $\sy{s} \in ( \sy{S} \cup \sy{R} )$
and $e \in E$ as $\sy{f}(\sy{s} \cdot e) = f(s \cdot e)$, $s \in \mu(\sy{s})$.

\begin{thm}[Automaton from Table]
    \label{th:table-aut}
From a closed, reduced and evidence compatible table one can construct a
deterministic symbolic automaton compatible with the concrete sample.
\end{thm}

\proof
The proof is similar to the concrete case. Let $\sy{T}=(\S, \sy{\S},
\sy{S}, \sy{R}, \psi, E, \sy{f}, \mu)$ be such a table, which is reduced and
closed and thus a function $g : \sy{R} \to \sy{S}$ such that $g(\sy{r}) =
\sy{s}$ if{f} $\sy{f}_{\sy{r}} = \sy{f}_{\sy{s}}$ is well defined.
The automaton derived from the table is then $\sy{\A}_{\sy{T}}=
(\S,\sy{\S},\psi, Q,\sy{\d}, q_0, F)$ where:
    \begin{itemize}
      \itemsep.5em
        \item $Q = \sy{S}$, $q_0 = \sy{\e}$
        \item $F = \{\sy{s} \in \sy{S} \mid \sy{f}_{\sy{s}}(\e)
                      = +\}$
        \item $\sy{\d} : Q \times \sy{\S} \to Q$ is defined as
                $ \sy{\d}(\sy{s},\a) =
                   \left\{
                     \begin{array}{l l} \sy{s} \cdot \a
                            &  \mbox{ when } \sy{s} \cdot \a \in \sy{S}\\
                        g(\sy{s} \cdot \a)
                            &  \mbox{ when } \sy{s} \cdot \a \in \sy{R}
                     \end{array}
                   \right.$
    \end{itemize}
By construction and like the $L^*$ algorithm,  $\sy{\A}_{\sy{T}}$ classifies
correctly the symbolic sample and, due to evidence compatibility, this holds
also for
the concrete sample. \qed


\section{Learning Languages over Ordered Alphabets}\label{S:SymboliLearningTotalOrder}

In this section we present a symbolic learning algorithm starting with an
intuitive verbal description. The algorithmic scheme is similar to the concrete
$L^{*}$ algorithm but differs in the treatment of counter-examples and the new
concept of evidence compatibility. Whenever the table is not closed, $\sy{S}\cup
\sy{R}$ is extended until closure. Then a conjectured automaton
$\sy{\A}_{\sy{T}}$ is constructed and an equivalence query is posed. If the
answer is positive we are done. Otherwise, the teacher provides a
counter-example leading to the extension of $\sy{S}\cup\sy{R}$ and/or  $E$. 
Whenever such an extension occurs, additional membership queries are
posed to fill the table. The table is always kept evidence compatible and
reduced except temporarily during the processing of counter-examples.

From now on we assume $\S$ to be a  \emph{totally ordered} alphabet with a minimal
element $a_0$  and restrict ourselves to symbolic automata where the concrete
semantics for every symbolic letter is an interval. In the case of a dense order like in $\R$, we assume the intervals to be left-closed and right-open. The order on the alphabet can  be extended naturally to a lexicographic order on $\S^{*}$. Our algorithm
also assumes that the teacher provides a counter-example of minimal length which
is  minimal with respect to the lexicographic order. This strong assumption
improves the performance of the algorithm and its relaxation is discussed in
Section~\ref{S:discussion}.

The rows of the observation table consist of symbolic words because we want to
group together all concrete letters and words that are assumed to induce the
same behavior in the automaton. New symbolic letters are introduced in two
occasions: when a new state is discovered or when a partition is modified due to
a counter-example. In both cases we set the concrete semantics $[\a]$ to the
largest possible subset of $\S$, given the current evidence (in the first case
it will be $\S$). As an evidence we always select the smallest possible $a \in
[\a]$ ($a_0$ when $[\a]=\S$).  The choice of the right evidences is a key
point for the performance of the algorithm as we want to  keep the concrete
sample as small as possible and avoid posing unnecessary queries. For infinite
concrete alphabets this choice of evidence guarantees termination.


\begin{algorithm}
  \caption{The symbolic algorithm}
  \label{alg:symbolic}
  \begin{algorithmic}[1]
    \Procedure{Symbolic}{}  \smallskip
    \State $learned=false$
    \State Initialize the table $\sy{T} = ( \Sigma, \sy{\Sigma}, \sy{S}, \sy{R},
\psi, E, \sy{f}, \mu)$
    \State \hspace{6pt} $\sy{\Sigma} = \{\sy{a}\}$; $ \psi_{\epsilon}(a) =
\sy{a}, \forall a \in \Sigma$
    \State \hspace{6pt} $\sy{S} = \{\sy{\epsilon}\}$; $\sy{R} = \{\sy{a}\}$; $ E
= \{\epsilon\}$
    \State \hspace{6pt} $\mu(\sy{a}) = \{a_0\}$ 
    \State \hspace{6pt} Ask MQ on $\epsilon$ and $a_0$ to fill $\sy{f}$
\smallskip

    \If {$\sy{T}$ is not closed}
        \State {\sc Close}
    \EndIf   \smallskip
    \Repeat
      \If {$EQ(\mathcal{A}_{\sy{T}})$}  \Comment{$\A_{\sy{T}}$ is correct}
          \State   $learned = true$
      \Else \Comment {A counter-example $w$ is provided}
	      \State $M=M\cup \{w\}$ 
	      \State {\sc Counter-ex($w$)}   \Comment{Process counter-example}
      \EndIf
    \Until {$learned$}
    \EndProcedure
  \end{algorithmic}
\end{algorithm}
\floatname{algorithm}{Procedure}

\begin{algorithm}
    \caption{Close the table}
    \label{alg:CloseTable}    \smallskip
    \begin{algorithmic}[1]
      \Procedure{Close}{} 
      \While {$\exists \sy{r} \in \sy{R}$ such that 
              $\forall \sy{s} \in \sy{S}$, $\sy{f}_{\sy{r}} \neq \sy{f}_{\sy{s}}$}
        \State {$\sy{S'} = \sy{S} \cup \{\sy{r}\}$} 
		\Comment {$\sy{r}$ becomes a new state}
        \State {$\sy{\S'} = \sy{\S} \cup \{\anew\}$}
        \State {$\psi' = \psi \cup \{\psi_{\sy{r}}\}$ with 
                $\psi_{\sy{r}}(a) = \anew$, $\forall a \in \S$}
        \State {$\sy{R'} = \left(\sy{R} - \{\sy{r}\} \right) \cup \{\sy{r} \cdot \anew\}$}
        \State {$\mu(\sy{r} \cdot \anew) = \mu(\sy{r}) \cdot a_0$} 
        \smallskip
    
        \State {Ask MQ for all words in
                	$ \{\mu (\sy{r} \cdot \anew) \cdot e:e \in E\}$}
	
        \State {$\sy{T} = ( \S, \sy{\S'}, \sy{S'}, \sy{R'}, \psi', E, \sy{f'}, \mu')$}
      \EndWhile
      \EndProcedure
    \end{algorithmic}
\end{algorithm}

\begin{algorithm}
    \caption{Process counter-example}
    \label{alg:TreatCex}    \smallskip
    \begin{algorithmic}[1]
      \Procedure{Counter-ex}{$w$} 
      \State {Find a factorization $w = u \cdot b \cdot v$, 
              $b \in \S$, $u,\,v \in \S^{*}$ such that }
      \State {\hspace{12pt} $\exists \sy{u} \in \sy{M_T},\,u \in \mu (\sy{u})$ 
              and $\forall \sy{u'} \in \sy{M_T},\,u \cdot b \notin \mu(\sy{u'})$ }
      \medskip

      \If {$\sy{u} \in \sy{S}$ } 	      \Comment {$\sy{u}$ is already a state}
          \State {Find $\a \in \sy{\S_u}$ such that $b \in [\a]$} \Comment {refine $[\a]$}
          \State {$\sy{\S'} = \sy{\S} \cup \{\anew\}$}
          \State {$\sy{R'} = \sy{R} \cup \{\sy{u}\cdot  \anew \}$}
          \State {$\mu(\sy{u} \cdot \anew) = \mu(\sy{u}) \cdot b $}
          \State {Ask MQ for all words in $\{ \mu(\sy{u} \cdot \anew) \cdot e:e\in E\}$}
          \State {$\psi'_{\sy{u}}(a) = \left\lbrace 
                  \begin{array}{ll}
                      \psi_{\sy{u}}(a)      & \text{  if } a \notin [\a] \\ 
                      \anew               & \text{  if } a \in [\a] 
					      \text{ and } a \geq b \\ 
                      \a                & \text{  otherwise }           
                  \end{array}  \right.$} 
          \smallskip
          \State {$\sy{T} =( \S, \sy{\S'}, $ $\sy{S}, \sy{R'}, \psi', E, \sy{f'}, \mu')$}
          \smallskip
      \Else \Comment {$\sy{u}$ is in the boundary}
	  \smallskip
	  \State {$\sy{S'} = \sy{S} \cup \{\sy{u}\}$} 
		  \Comment {and becomes a state}
          \If {$b = a_0$}     
              \State {$\sy{\S'} = \sy{\S} \cup \{\anew\}$}
	      \State {$\psi' = \psi \cup \{\psi_{\sy{u}}\}$, 
                      with $\psi_{\sy{u}}(a) = \anew, \forall a \in \S$}
              \State {$\sy{R'} = (\sy{R}- \{\sy{u}\}) \cup \{\sy{u} \cdot \anew \}$}
              \State {$E' = E \cup \{\text{suffixes of } b\cdot v \}$}
              \State {$\mu(\sy{u} \cdot \anew) = \mu(\sy{u}) \cdot a_0$}
              \State {Ask MQ for all words in $\{ \mu(\sy{u} \cdot \anew) \cdot e:e \in E'\}$}
          \Else                
              \State {$\sy{\S'} = \sy{\S} \cup \{\anew,\anew'\}$}
              \State {$\psi' = \psi \cup \{\psi_{\sy{u}}\}$, with
                      $\psi_{\sy{u}}(a) = \left\lbrace 
                          \begin{array}{ll}	
                              \anew'  & \text{ if } a \geq b \\
			      \anew & \text{ otherwise}
		          \end{array} \right.$} 
			  \smallskip
              \State {$\sy{R'} = (\sy{R}- \{\sy{u}\}) \cup \{\sy{u} \cdot \anew, \sy{u} \cdot \anew' \}$}
              \State {$E' = E \cup \{\text{suffixes of } b\cdot v \}$}
              \State {$\mu(\sy{u} \cdot \anew) = \mu(\sy{u}) \cdot a_0$; 
                      $\mu(\sy{u} \cdot \anew') = \mu(\sy{u}) \cdot b$}
              \State {Ask MQ for all words in $\{\left(\mu(\sy{u} \cdot \anew) \cup
              		 \mu(\sy{u} \cdot \anew')\right) \cdot e:e\in E'\}$}
          \EndIf
          \State {$\sy{T} =( \S, \sy{\S'},\sy{S'}, \sy{R'}, \psi', E', \sy{f'}, \mu')$}           
      \EndIf
      \If {$\sy{T}$ is not closed}
          \State {\sc close}
      \EndIf   \smallskip
      \EndProcedure
    \end{algorithmic}
\end{algorithm}
\floatname{algorithm}{Algorithm}

The initial symbolic table is $\sy{T} = ( \S, \sy{\S}, $ $\sy{S}, \sy{R}, \psi,
E, \sy{f}, \mu)$, where $\sy{\S} = \{\sy{a_0}\}$, $[\a_0] = \S$, $\sy{S} = \{
\sy{\e} \} $, $\sy{R} = \{\sy{a_0}\}$, $E = \{\e \}$, and
$\mu(\sy{a_0})=\{a_0\}$. The table is filled by membership queries concerning
$\e$ and $a_0$. Whenever $\sy{T}$ is not closed, there is some $\sy{r} \in
\sy{R}$ such that $\sy{f}_{\sy{r}} \neq \sy{f}_{\sy{s}}$ for every $\sy{s} \in \sy{S}$. To
close the table we move  $\sy{r}$ from $\sy{R}$ to $\sy{S}$, recognizing it as a
new state, and checking the behavior of its continuation. To this end we add to
$\sy{R}$ the word $\sy{r'} = \sy{r} \cdot \a$, where $\a$ is a new symbolic
letter with $[\a] = \S$.
We extend the evidence function by letting $\mu(\sy{r'})=\mu(\sy{r}) \cdot a_0$,
assuming that all elements of $\S$ behave as $a_0$ from $\sy{r}$.
Once $\sy{T}$ is closed we construct a hypothesis automaton as described in
the proof of Theorem~\ref{th:table-aut}.

When a counter-example $w$ is presented, it is of course not part of the
concrete sample. A miss-classified word in the conjectured automaton means that
somewhere a wrong transition is taken. Hence  $w$ admits a factorization $w=u
\cdot b \cdot v$ where $u \in \S^{*}$ and $b \in \S$ is where the first wrong
transition is taken. Obviously we do not know $u$ and $b$ in advance but know
that this
happens in the following two cases. Either $b$ leads to an undiscovered state in
the automaton of the target language, or letter $b$ does not belong to the interval
it was assumed to belong in the conjectured automaton. The latter case happens
only when $b$ does not belong to the evidence function. Since counter-example $w$
is minimal, it admits a factorization $w=u \cdot b \cdot v$, where $u$ is the
largest prefix of $w$ such that $u \in \mu(\sy{u})$ for some $\sy{u}\in
\sy{S}\cup \sy{R}$ but $s\cdot b \notin \mu(\sy{u}')$ for any word $\sy{u}'$ in
the symbolic sample. We consider two cases, $\sy{u}\in \sy{S}$ and $\sy{u}\in\sy{R}$.

In the first case, when $\sy{u}\in \sy{S}$, $\sy{u}$ is already a state in the hypothesis
but $b$ indicates that the partition boundariues are not correctly defined and
need refinement. That is, $u\cdot b $ was wrongly considered to be part of $[\sy{u}\cdot \a]$
for some $\a \in \sy{\S}_{\sy{u}}$, and thus $b$ was wrongly considered to be
part of $[\a]$.
Due to minimality, all letters in $[\a]$ less than letter $b$ behave like $\mu(\a)$.
We assume that  all remaining letters in $[\a]$ behave like $b$ and
map them to a new symbol $\anew$ that we add to $\sy{\S}_{\sy{u}}$.
We then update $\psi_{\sy{u}}$ such that $\psi'_{\sy{u}}(a) = \anew$ for all $a \in [\a], a\geq b$, and
$\psi'_{\sy{u}}(a) = \psi_{\sy{u}}(a)$, otherwise. The evidence function is updated by
letting $\mu(\sy{u} \cdot \anew) = \mu(\sy{u}) \cdot b$ and $\sy{u}\cdot\anew$ is added to $\sy{R}$.

In the second case, the symbolic word $\sy{u}$ is part of the boundary.
From the counterexample we deduce that $\sy{u}$ is not equivalent to any of the
existing states in the hypothesis and should form a new state.
Specifically, we find the prefix $\sy{s}$ that was considered to be equivalent
to $\sy{u}$, that is $g(\sy{u})=\sy{s}\in \sy{S}$. Since the table is reduced
$\sy{f}_{\sy{u}}\not=\sy{f}_{\sy{s'}}$ for any other $\sy{s'}\in \sy{S}$.
Because $w$ is the shortest counter-example, the classification of $\sy{s}\cdot
b \cdot v$ in the automaton is correct (otherwise $s\cdot b \cdot v$, for some
$s\in [\sy{s}]$ would constitute a shorter counter-example) and different from
that of $u\cdot b \cdot v$. We conclude that $\sy{u}$ is a new state, which is added to
$\sy{S}$. To distinguish between $\sy{u}$ and $\sy{s}$ we add to $E$ the word
$b \cdot v$, possibly with some of its suffixes (see \cite{BergR04}
for a more detailed discussion of counter-example processing).

As $\sy{u}$ is a new state we need to add its continuations to $R$. We
distinguish two subcases depending on $b$.
If $b = a_0$, the smallest element of $\S$, then a new symbolic letter $\anew$
is added to $\sy{\S}$, with $[\anew] = \S$ and $\mu(\sy{u} \cdot \anew) =
\mu(\sy{u}) \cdot a_0$, and the symbolic word $\sy{u} \cdot \anew$ is added to
$\sy{R}$.
If $b\not = a_0$ then \emph{two} new symbolic letters, $\anew$ and
$\anew'$, are added to $\sy{\S}$ with $[\anew] = \{a : a < b\}$, $[\anew'] =
\{a  : a \geq b\}$, $\mu(\sy{u} \cdot \anew) = \mu(\sy{u}) \cdot a_0$ and
$\mu(\sy{u} \cdot \anew') = \mu(\sy{u}) \cdot b$. The words $\sy{u}\cdot
\anew$ and $\sy{u}\cdot  \anew'$ are added to $\sy{R}$.


A detailed description of the algorithm is given in Algorithm~\ref{alg:symbolic}
and its major procedures, table closing and counter-example treatment are described
in Procedures~\ref{alg:CloseTable} and \ref{alg:TreatCex} respectively.
A statement of the form $\sy{\S}'=\sy{\S}\cup\{\a\}$ indicates the introduction
of a new symbolic letter $\a\not\in \sy{\S}$. We use $MQ$ and $EQ$ as shorthands
for membership and equivalence queries, respectively. In the following we illustrate the symbolic algorithm as applied to a
language over an infinite alphabet.

%
%
 
\begin{figure}[ht]
\begin{tabular}{|c|}
  \hline
    \begin{tabular}{|@{ }r@{ }|@{ }c@{ }|}
     \multicolumn{2}{c}{$\sy{T}_0$} \\[4pt]
        \hline
         & $\epsilon$ \\
        \hline
        $\epsilon$ & - \\
        \hline
        $\a_0$     & + \\
        \hline
    \end{tabular}
    \hfill
    \begin{tabular}{|@{ }r@{ }|@{ }c@{ }|}
    \multicolumn{2}{c}{$\sy{T}_1$} \\[4pt]
        \hline
         & $\epsilon$ \\
        \hline
        $\epsilon$  & - \\
        $\a_0$      & + \\
        \hline
        $\a_0 \cdot \a_1$ & + \\
        \hline
    \end{tabular}
    \hfill
    \begin{tabular}{|@{ }r@{ }|@{ }c@{ }|}
    \multicolumn{2}{c}{$\sy{T}_2$} \\[4pt]
        \hline
         & $\epsilon$ \\
        \hline
        $\epsilon$  & - \\
        $\a_0$      & + \\
        \hline
        $\a_0 \cdot \a_1$ & + \\
        $\a_2$      & - \\
        \hline
    \end{tabular}
    \hfill
    \begin{tabular}{|@{ }r@{ }|@{ }c@{ }|}
    \multicolumn{2}{c}{$\sy{T}_3$} \\[4pt]
        \hline
         & $\epsilon$ \\
        \hline
        $\epsilon$  & - \\
        $\a_0$      & + \\
        \hline
        $\a_0 \cdot \a_1$ & + \\
        $\a_2$      & - \\
        $\a_0 \cdot \a_3$ & - \\
        \hline
    \end{tabular}
    \hfill
    \begin{tabular}{|@{ }r@{ }|@{ }cc@{ }|}
    \multicolumn{3}{c}{$\sy{T}_4$} \\[4pt]
        \hline
         & $\epsilon$ & $0$\\
        \hline
        $\epsilon$  & - & + \\
        $\a_0$      & + & + \\
        $\a_2$      & - & - \\
        \hline
        $\a_0 \cdot \a_1$ & + & - \\
        $\a_0 \cdot \a_3$ & - & - \\
        $\a_2 \cdot \a_4$ & - & + \\
        \hline
    \end{tabular}  \\
     \\\hline
    \begin{tabular}{|@{ }r@{ }|@{ }cc@{ }|}
    \multicolumn{3}{c}{$\sy{T}_5$} \\[4pt]
        \hline
         & $\epsilon$ & $0$\\
        \hline
        $\epsilon$  & - & + \\
        $\a_0$      & + & + \\
        $\a_2$      & - & - \\
        $\a_0 \cdot \a_1$ & + & - \\
        \hline
        $\a_0 \cdot \a_3$ & - & - \\
        $\a_2 \cdot \a_4$ & - & + \\
        $\a_0 \cdot \a_1 \cdot \a_5$ & - & - \\
        \hline
    \end{tabular}  
    \hfill
    \begin{tabular}{|@{ }r@{ }|@{ }cc@{ }|}
    \multicolumn{3}{c}{$\sy{T}_6$} \\[4pt]
        \hline
         & $\epsilon$ & $0$\\
        \hline
        $\epsilon$  & - & + \\
        $\a_0$      & + & + \\
        $\a_2$      & - & - \\
        $\a_0 \cdot \a_1$ & + & - \\
        \hline
        $\a_0 \cdot \a_3$ & - & - \\
        $\a_2 \cdot \a_4$ & - & + \\
        $\a_0 \cdot \a_1 \cdot \a_5$ & - & - \\
        $\a_2 \cdot \a_6$ & + & - \\
        \hline
    \end{tabular}  
    \hfill
    \begin{tabular}{|@{ }r@{ }|@{ }cc@{ }|}
    \multicolumn{3}{c}{$\sy{T}_7$} \\[4pt]
        \hline
         & $\epsilon$ & $0$\\
        \hline
        $\epsilon$  & - & + \\
        $\a_0$      & + & + \\
        $\a_2$      & - & - \\
        $\a_0 \cdot \a_1$ & + & - \\
        \hline
        $\a_0 \cdot \a_3$ & - & - \\
        $\a_2 \cdot \a_4$ & - & + \\
        $\a_0 \cdot \a_1 \cdot \a_5$ & - & - \\
        $\a_2 \cdot \a_6$ & + & - \\
        $\a_2 \cdot \a_7$ & - & - \\
        \hline
    \end{tabular} 
    \hfill
    \begin{tabular}{|@{ }r@{ }|@{ }cc@{ }|}
    \multicolumn{3}{c}{$\sy{T}_8$} \\[4pt]
        \hline
         & $\epsilon$ & $0$\\
        \hline
        $\epsilon$  & - & + \\
        $\a_0$      & + & + \\
        $\a_2$      & - & - \\
        $\a_0 \cdot \a_1$ & + & - \\
        \hline
        $\a_0 \cdot \a_3$ & - & - \\
        $\a_2 \cdot \a_4$ & - & + \\
        $\a_0 \cdot \a_1 \cdot \a_5$ & - & - \\
        $\a_2 \cdot \a_6$ & + & - \\
        $\a_2 \cdot \a_7$ & - & - \\
        $\a_2 \cdot \a_8$ & + & + \\
        \hline
    \end{tabular}  \\
     \\\hline
\end{tabular}
    \smallskip
    \caption{Observation tables for Example~\ref{ex:SLex2}.}
    \label{fig:SLex2_tables}
\end{figure}

\begin{figure}[ht]
\resizebox{\textwidth}{!}{\begin{tabular}{| c | c | c |}
\hline
   $\sy{\A}_{1}$ & $\sy{\A}_{2}$ & $\sy{\A}_{3}$ \\
   \scalebox{1}{\input{./Figures/Ex_SL2/SLex2_psi1}} &
   \scalebox{1}{\input{./Figures/Ex_SL2/SLex2_psi2}} &
   \scalebox{1}{\input{./Figures/Ex_SL2/SLex2_psi3}} \\
   \scalebox{1.3}{\input{./Figures/Ex_SL2/SLex2_h1}} &
   \scalebox{1.3}{\input{./Figures/Ex_SL2/SLex2_h2}} &
   \scalebox{1.3}{\input{./Figures/Ex_SL2/SLex2_h3}}  \\
\hline
\end{tabular}}\par
\resizebox{\textwidth}{!}{\begin{tabular}{| c | c |}
   $\sy{\A}_{5}$ & $\sy{\A}_{6}$ \\
   \scalebox{.9}{\input{./Figures/Ex_SL2/SLex2_psi5}} &
   \scalebox{.9}{\input{./Figures/Ex_SL2/SLex2_psi6}} \\
   \scalebox{1.1}{\input{./Figures/Ex_SL2/SLex2_h5}}  &
   \scalebox{1.1}{\input{./Figures/Ex_SL2/SLex2_h6}} \\
\hline
   $\sy{\A}_{7}$ & $\sy{\A}_{8}$ \\
   \scalebox{.9}{\input{./Figures/Ex_SL2/SLex2_psi7}} &
   \scalebox{.9}{\input{./Figures/Ex_SL2/SLex2_psi8}} \\
   \scalebox{1.1}{\input{./Figures/Ex_SL2/SLex2_h7}} &
   \scalebox{1.1}{\input{./Figures/Ex_SL2/SLex2_h8}} \\
\hline
\end{tabular}}
    \caption{Hypotheses and $\S$-semantics for Example~\ref{ex:SLex2}}
    \label{fig:SLex2_hypotheses}
\end{figure}

\begin{exa}\label{ex:SLex2}
Let  $\S  = [0,100) \subset \R$ with the usual order and let  $L \subseteq
\S^{*}$ be a target language. Fig.~\ref{fig:SLex2_tables} shows the evolution of the symbolic observation
tables and Fig.~\ref{fig:SLex2_hypotheses} depicts the corresponding automata and the concrete semantics of the symbolic alphabets.

We initialize the table with $\sy{S}=\{\sy{\e}\}$, $\sy{R}=\{\a_0\}$,
$\mu(\a_0)=\{0\}$ and $E=\{\e\}$ and ask membership queries for $\e$ (rejected)
and $0$ (accepted). The obtained table, $\sy{T}_0$ is not closed so we move $\a_0$ to
$\sy{S}$, introduce $\sy{\S}_{\a_0} = \{\a_1\}$, where $\a_1$ is a new symbol, and add $\a_0 \cdot \a_1$ to $R$ with
$\mu(\a_0 \cdot \a_1)=0 \cdot 0$. Asking membership queries we obtain the closed table
$\sy{T}_1$ and its automaton $\sy{\A}_1$. We pose an equivalence query and obtain $(50,-)$
as a (minimal) counter-example which implies that all words smaller than $50$
are correctly classified. We add a new symbol $\a_2$ to $\sy{\S}_{\sy{\e}}$ and
redefine the concrete semantics to $[\a_0]=\{a<50\}$ and $[\a_2]=\{a\geq 50\}$. As
evidence we select the smallest possible letter, $\mu(\a_2)=50$, ask membership queries
to obtain the closed table $\sy{T}_2$ and automaton $\sy{\A}_2$.

For this hypothesis we get a counter-example $(0 \cdot 30,-)$ whose prefix $0$ is
already in the sample, hence the misclassification occurs in the second
transition. We refine the alphabet partition for state $\a_0$ by introducing a
new symbol $\a_3$ and letting $[\a_1]=\{a<30\}$ and $[\a_3]=\{a\geq 30\}$. Table $\sy{T}_3$
is closed but automaton $\sy{\A}_3$ is still incorrect and a counter-example
$(50\cdot 0,-)$ is provided. The prefix $50$ belongs to the evidence of $\a_2$
and is moved from the boundary to become a new state and its successor $\a_2 \cdot
\a_4$, for a new symbol $\a_4$, is added to $\sy{R}$. To distinguish $\a_2$ from
$\sy{\e}$, the suffix $0$ of the counter-example is added to $E$ resulting in
$\sy{T}_4$ which is not closed. The newly discovered state $\a_0 \cdot \a_1$ is added to
$\sy{S}$, the filled table $\sy{T}_5$ is closed and the conjectured automaton $\sy{A}_5$
has two additional states.

Subsequent equivalence queries result counter-examples $(50\cdot 20,+)$,
$(50\cdot 80,-)$ and $(50\cdot 50 \cdot 0,+)$ which are used to refine the
alphabet partition at state $\a_2$ and modify its outgoing transitions
progressively as seen in automata $\sy{\A}_6$, $\sy{\A}_7$ and $\sy{\A}_8$, respectively.
Automaton $\sy{\A}_8$ accepts the target language and the algorithm terminates. \qed
\end{exa}

\ni Note that for the language in Example~\ref{ex:Lstar},
the symbolic algorithm needs around $30$ queries instead of the $80$ queries
required by $L^*$. If we choose to learn a language as the one
described in Example~\ref{ex:SLex2}, restricting the concrete alphabet to the
finite alphabet $\S = \{1,\dots, 100\}$, then $L^*$ requires around $1000$
queries compared to $17$ queries required by our symbolic algorithm.
As we shall see in Section~\ref{S:complexity}, the complexity of the symbolic
algorithm does not depend on the size of the concrete alphabet, only on the
number of transitions.

\section{Learning Languages over Partially-ordered Alphabets}\label{S:PartialOrderedAlphabets}

In this section we sketch the extension of the results of this paper to
partially-ordered alphabets of the form $\S=X^d$ where $X$ is a totally-ordered
set such as an interval $[0,k)\se\R$.  Letters of $\S$ are $d$-tuples of the form
$\vx = \left(x_1, \dots, x_d\right)$ and the minimal element is $\vzero =
\left(0, \dots, 0\right)$.
The usual partial order on this set is defined as $\vx \leq \vy$ if
and only if $x_i \leq y_i$ for all $i = 1,\dots,d$. When $\vx  \leq \vy$ and
$x_i \neq y_i$ for some $i$ the inequality is strict, denoted by $\vx<\vy$, and
we say then that $\vx$ \emph{dominates} $\vy$. Two elements are
\emph{incomparable}, denoted by $\vx || \vy$, if $x_i<y_i$ and $x_j>y_j$ for
some $i$ and $j$.

For partially-ordered sets, a natural extension of the partition of an
ordered set into intervals is a \emph{monotone} partition, where for each
partition block $P$ there are no three points such that $\vx<\vy<\vz$,
$\vx,\vz\in P$, and $ \vy\not \in P$. We define in the following such
partitions represented by a finite set of points.

A \emph{forward cone} $B^+(\vx) \subset \S$ is the set of all
points dominated by a point $\vx\in\S$ (see Fig.~\ref{Fig:forwardAndBackwardCones}).
Let $F=\{\vx_1,\ldots,\vx_l\}$ be a set of points, then $B^+(F)=B^+(\vx_1) \cup
\ldots \cup B^+(\vx_l)$ as shown in Fig.~\ref{Fig:unionOfForwardCones}.
From a family of sets of points $\mathcal{F}=\{F_0,\ldots,F_{m-1}\}$,
such that $F_0=\{\vzero\}$ satisfying for every $i$:
1) $\forall \vy\in F_i,~\exists \vx \in F_{i-1}$ such that $\vx<\vy$, and
2) $\forall \vy\in F_i,~\forall \vx \in F_{i-1}, ~\vy \not < \vx$,
we can define a monotone partition of the form $\mathcal{P} = \{P_1, \dots, P_{m-1}\}$,
where $P_i = B^+(F_{i-1})-B^+(F_i)$, see  Fig.~\ref{Fig:aPartition}.

\begin{figure}
\begin{subfigure}[t]{.3\textwidth}
\centering
\scalebox{.55}{\input{./Figures/cones/forwardAndBackwardCones}}
\caption{Backward and forward cone for $\vx$}
\label{Fig:forwardAndBackwardCones}
\end{subfigure}
\begin{subfigure}[t]{.3\textwidth}
\centering
\scalebox{.55}{\input{./Figures/cones/unionOfForwardCones}}
\caption{Union of cones}
\label{Fig:unionOfForwardCones}
\end{subfigure}
\begin{subfigure}[t]{.3\textwidth}
\centering
\scalebox{.55}{\input{./Figures/cones/partition}}
\caption{An alphabet partition}
\label{Fig:aPartition}
\end{subfigure}

\caption{ }
\label{Fig:example_automaton2tree}
\end{figure}

\begin{figure}

  \begin{subfigure}[t]{.4\textwidth}
      \centering
      \scalebox{.55}{\input{./Figures/cones/SL_refine_partition_2}}
      \caption{ } \label{fig:part_ref_a}
  \end{subfigure}
  \begin{subfigure}[t]{.4\textwidth}
      \centering
      \scalebox{.55}{\input{./Figures/cones/SL_refine_partition_3}}
      \caption{ } \label{fig:part_ref_b}
  \end{subfigure}

   \caption{Modifying the alphabet partition for state $\sy{u}$ after receiving $u\cdot b \cdot v$ as counter-example. 
   Letters above $b$ are moved from $[ \a]$ to $[ \a']$. }
\end{figure}

A subset $P$ of $\S$, as defined above, may have several mutually-incomparable minimal
elements, none of which being dominated by any other element of $P$.
One can thus apply the symbolic learning algorithm but without the presence of
unique minimal evidence and minimal counter-example.
For this reason a symbolic word may have more than one evidence. Evidence
compatibility is preserved though due to the nature of the partition.

The teacher is assumed to return a counter-example chosen from a set of
incomparable minimal counter-examples. Like in the algorithm for totally ordered alphabet,
every counter-example either discovers a new state or refines a partition.
The learning algorithm for partially-ordered alphabets is similar to Algorithm~\ref{alg:symbolic}
and can be applied with only a minor modification in the treatment of the counterexamples
and specifically in the refinement procedure.
Lines 6-8 of Procedure~\ref{alg:TreatCex} should be ignored in the case where there
exists a symbolic letter $\a'$, as illustrated in Fig.~\ref{fig:part_ref_a}, such
that $\sy{f}(\sy{u}\cdot b \cdot e ) = \sy{f}(\sy{u} \cdot \a'  \cdot e )$ for all $e \in E$.
In such a case, function $\psi$ is updated as in line 9 by replacing $\anew$ by $\a'$ and $b$
should be added to $\mu(\a')$. In Fig.~\ref{fig:part_ref_b}, one can see
the partition after refinement, where all letters above $b$ have been moved from $[\a]$ to $[\a']$.


%
%
 
\begin{figure}[ht]
    \scalebox{.9}{

\begin{tabular}{|cccc|}
  \hline
    \begin{tabular}{|@{ }r@{ }|@{ }c@{ }|}
     \multicolumn{2}{c}{$\sy{T}_0$} \\[4pt]
        \hline
         & $\epsilon$ \\
        \hline
        $\epsilon$  & - \\
        $\a_0$      & + \\
        \hline
        $\a_0 \cdot \a_1$ & + \\
        \hline
    \end{tabular}
    &
    \begin{tabular}{|@{ }r@{ }|@{ }c@{ }|}
     \multicolumn{2}{c}{$\sy{T}_{1-3}$} \\[4pt]
        \hline
         & $\epsilon$ \\
        \hline
        $\epsilon$  & - \\
        $\a_0$      & + \\
        \hline
        $\a_0 \cdot \a_1$ & + \\
        $\a_2$      & - \\
        \hline
    \end{tabular}
    &
    \begin{tabular}{|@{ }r@{ }|@{ }c@{ }|}
    \multicolumn{2}{c}{$\sy{T}_{4-7}$} \\[4pt]
        \hline
         & $\epsilon$ \\
        \hline
        $\epsilon$  & - \\
        $\a_0$      & + \\
        \hline
        $\a_0 \cdot \a_1$ & + \\
        $\a_2$      & - \\
        $\a_0 \cdot \a_3$ & - \\
        \hline
    \end{tabular}
    &
    \begin{tabular}{|@{ }r@{ }|@{ }cc@{ }|}
    \multicolumn{3}{c}{$\sy{T}_8$} \\[4pt]
        \hline
         & $\epsilon$ & {\small $\binom {0} {0}$} \\
        \hline
        $\epsilon$  & - & + \\
        $\a_0$      & + & + \\
        $\a_2$      & - & - \\
        \hline
        $\a_0 \cdot \a_1$ & + & - \\
        $\a_0 \cdot \a_3$ & - & - \\
        $\a_2 \cdot \a_4$ & - & + \\
        \hline
    \end{tabular}\\[2cm]
    \hline
    \begin{tabular}{|@{ }r@{ }|@{ }cc@{ }|}
    \multicolumn{3}{c}{$\sy{T}_9$} \\[4pt]
        \hline
         & $\epsilon$ & {\small $\binom {0} {0}$} \\
        \hline
        $\epsilon$  & - & + \\
        $\a_0$      & + & + \\
        $\a_2$      & - & - \\
        $\a_0 \cdot \a_1$ & + & - \\
        \hline
        $\a_0 \cdot \a_3$ & - & - \\
        $\a_2 \cdot \a_4$ & - & + \\
        $\a_0 \cdot \a_1 \cdot \a_5$ & - & - \\
        \hline
    \end{tabular}
    &
    \begin{tabular}{|@{ }r@{ }|@{ }cc@{ }|}
    \multicolumn{3}{c}{$\sy{T}_{10-11}$} \\[4pt]
        \hline
         & $\epsilon$ & {\small $\binom {0} {0}$} \\
        \hline
        $\epsilon$  & - & + \\
        $\a_0$      & + & + \\
        $\a_2$      & - & - \\
        $\a_0 \cdot \a_1$ & + & - \\
        \hline
        $\a_0 \cdot \a_3$ & - & - \\
        $\a_2 \cdot \a_4$ & - & + \\
        $\a_0 \cdot \a_1 \cdot \a_5$ & - & - \\
        $\a_2 \cdot \a_6$ & + & - \\
        \hline
    \end{tabular}
    &
    \begin{tabular}{|@{ }r@{ }|@{ }cc@{ }|}
    \multicolumn{3}{c}{$\sy{T}_{12-15}$} \\[4pt]
        \hline
         & $\epsilon$ & {\small $\binom {0} {0}$} \\
        \hline
        $\epsilon$  & - & + \\
        $\a_0$      & + & + \\
        $\a_2$      & - & - \\
        $\a_0 \cdot \a_1$ & + & - \\
        \hline
        $\a_0 \cdot \a_3$ & - & - \\
        $\a_2 \cdot \a_4$ & - & + \\
        $\a_0 \cdot \a_1 \cdot \a_5$ & - & - \\
        $\a_2 \cdot \a_6$ & + & - \\
        $\a_2 \cdot \a_7$ & - & - \\
        \hline
    \end{tabular} 
    &
    \begin{tabular}{|@{ }r@{ }|@{ }cc@{ }|}
    \multicolumn{3}{c}{$\sy{T}_{16-18}$} \\[4pt]
        \hline
         & $\epsilon$ & {\small $\binom {0} {0}$} \\
        \hline
        $\epsilon$  & - & + \\
        $\a_0$      & + & + \\
        $\a_2$      & - & - \\
        $\a_0 \cdot \a_1$ & + & - \\
        \hline
        $\a_0 \cdot \a_3$ & - & - \\
        $\a_2 \cdot \a_4$ & - & + \\
        $\a_0 \cdot \a_1 \cdot \a_5$ & - & - \\
        $\a_2 \cdot \a_6$ & + & - \\
        $\a_2 \cdot \a_7$ & - & - \\
        $\a_2 \cdot \a_8$ & + & + \\
        \hline
    \end{tabular} \\
     & & & \\\hline
\end{tabular}}
    \smallskip
    \caption{Observation tables for Example~\ref{ex:SLex3} }
    \label{fig:SLex3_tables}
\end{figure}

\begin{figure}[ht]
    \centering
\def\H#1 { \input{./Figures/Ex_SL3/SLex3_h#1} }
\scalebox{1}{
    \begin{tabular}{| c | c | c |}
\hline
   $\sy{A}_0$ & $\sy{A}_{1-3}$ & $\sy{A}_{4-7}$ \\
     \input{./Figures/Ex_SL3/SLex3_h0}  &  \input{./Figures/Ex_SL3/SLex3_h1}  &  \input{./Figures/Ex_SL3/SLex3_h4}  \\
\hline
\end{tabular}} \par
\scalebox{1.12}{
\begin{tabular}{| c | c |}
   $\sy{A}_9$ & $\sy{A}_{10-11}$  \\
    \input{./Figures/Ex_SL3/SLex3_h8}   &  \input{./Figures/Ex_SL3/SLex3_h9}   \\
\hline
\end{tabular}}\par
\scalebox{1.12}{
\begin{tabular}{| c | c |}
   $\sy{A}_{12-15}$ & $\sy{A}_{16-18}$ \\
    \input{./Figures/Ex_SL3/SLex3_h11}        &  \input{./Figures/Ex_SL3/SLex3_h15}        \\
\hline
\end{tabular}}
    \caption{Hypothesis automata for Example~\ref{ex:SLex3} }
    \label{fig:SLex3_hypotheses_automata}
\end{figure}

\begin{figure}[ht]
    \centering
\def\SLp#1 {\scalebox{.5}{\input{./Figures/Ex_SL3/SLex3_all_psi/SLex3_psi#1} }}
\begin{tabular}{r l}
  $\psi_0$ & \scalebox{.5}{\input{./Figures/Ex_SL3/SLex3_all_psi/SLex3_psi0_a0} } \scalebox{.5}{\input{./Figures/Ex_SL3/SLex3_all_psi/SLex3_psi1_a1} } \\
  $\psi_1$ & \scalebox{.5}{\input{./Figures/Ex_SL3/SLex3_all_psi/SLex3_psi0_a0a2_a} } \scalebox{.5}{\input{./Figures/Ex_SL3/SLex3_all_psi/SLex3_psi1_a1} } \\
  $\psi_2$ & \scalebox{.5}{\input{./Figures/Ex_SL3/SLex3_all_psi/SLex3_psi0_a0a2_b} } \scalebox{.5}{\input{./Figures/Ex_SL3/SLex3_all_psi/SLex3_psi1_a1} } \\
  $\psi_3$ & \scalebox{.5}{\input{./Figures/Ex_SL3/SLex3_all_psi/SLex3_psi0_a0a2} } \scalebox{.5}{\input{./Figures/Ex_SL3/SLex3_all_psi/SLex3_psi1_a1} } \\
\end{tabular}
\begin{tabular}{r l}
  $\psi_4$ & \scalebox{.5}{\input{./Figures/Ex_SL3/SLex3_all_psi/SLex3_psi0_a0a2} } \scalebox{.5}{\input{./Figures/Ex_SL3/SLex3_all_psi/SLex3_psi1_a1a3_a} } \\
  $\psi_5$ & \scalebox{.5}{\input{./Figures/Ex_SL3/SLex3_all_psi/SLex3_psi0_a0a2} } \scalebox{.5}{\input{./Figures/Ex_SL3/SLex3_all_psi/SLex3_psi1_a1a3_b} } \\
  $\psi_6$ & \scalebox{.5}{\input{./Figures/Ex_SL3/SLex3_all_psi/SLex3_psi0_a0a2} } \scalebox{.5}{\input{./Figures/Ex_SL3/SLex3_all_psi/SLex3_psi1_a1a3_c} } \\
  $\psi_7$ & \scalebox{.5}{\input{./Figures/Ex_SL3/SLex3_all_psi/SLex3_psi0_a0a2} } \scalebox{.5}{\input{./Figures/Ex_SL3/SLex3_all_psi/SLex3_psi1_a1a3} } \\
\end{tabular}
\caption{Alphabet partition for Example~\ref{ex:SLex3} (part 1)}
    \label{fig:SLex3_psi2}
\end{figure}

 \begin{figure}
 \ContinuedFloat
    \centering
\def\SLp#1 {\scalebox{.5}{\input{./Figures/Ex_SL3/SLex3_all_psi/SLex3_psi#1} }}
\begin{tabular}{rl}
  $\psi_8$ & \scalebox{.5}{\input{./Figures/Ex_SL3/SLex3_all_psi/SLex3_psi0_a0a2} } \scalebox{.5}{\input{./Figures/Ex_SL3/SLex3_all_psi/SLex3_psi1_a1a3} } \scalebox{.5}{\input{./Figures/Ex_SL3/SLex3_all_psi/SLex3_psi2_a4} }  \\
  $\psi_9$ & \scalebox{.5}{\input{./Figures/Ex_SL3/SLex3_all_psi/SLex3_psi0_a0a2} } \scalebox{.5}{\input{./Figures/Ex_SL3/SLex3_all_psi/SLex3_psi1_a1a3} } \scalebox{.5}{\input{./Figures/Ex_SL3/SLex3_all_psi/SLex3_psi2_a4} } \scalebox{.5}{\input{./Figures/Ex_SL3/SLex3_all_psi/SLex3_psi3_a5} } \\
  & \multicolumn{1}{c}{\vdots}   \\
  $\psi_{11}$ & \scalebox{.5}{\input{./Figures/Ex_SL3/SLex3_all_psi/SLex3_psi0_a0a2} } \scalebox{.5}{\input{./Figures/Ex_SL3/SLex3_all_psi/SLex3_psi1_a1a3} } \scalebox{.5}{\input{./Figures/Ex_SL3/SLex3_all_psi/SLex3_psi2_a4a6} } \scalebox{.5}{\input{./Figures/Ex_SL3/SLex3_all_psi/SLex3_psi3_a5} } \\
  & \multicolumn{1}{c}{\vdots}   \\
  $\psi_{15}$ & \scalebox{.5}{\input{./Figures/Ex_SL3/SLex3_all_psi/SLex3_psi0_a0a2} } \scalebox{.5}{\input{./Figures/Ex_SL3/SLex3_all_psi/SLex3_psi1_a1a3} } \scalebox{.5}{\input{./Figures/Ex_SL3/SLex3_all_psi/SLex3_psi2_a4a6a7} } \scalebox{.5}{\input{./Figures/Ex_SL3/SLex3_all_psi/SLex3_psi3_a5} } \\
  & \multicolumn{1}{c}{\vdots}   \\
  $\psi_{18}$ & \scalebox{.5}{\input{./Figures/Ex_SL3/SLex3_all_psi/SLex3_psi0_a0a2} } \scalebox{.5}{\input{./Figures/Ex_SL3/SLex3_all_psi/SLex3_psi1_a1a3} } \scalebox{.5}{\input{./Figures/Ex_SL3/SLex3_all_psi/SLex3_psi2_a4a6a7a8} } \scalebox{.5}{\input{./Figures/Ex_SL3/SLex3_all_psi/SLex3_psi3_a5} } \\
\end{tabular}

\caption{Alphabet partition for Example~\ref{ex:SLex3} (part 2)}
    \label{fig:SLex3_psi}
\end{figure}

\begin{exa}\label{ex:SLex3}
Let us illustrate the learning process for a target language $L$ defined over $\S = [0,100]^2$. 
All tables, hypotheses automata and alphabet partitions for this example are
shown in Figures \ref{fig:SLex3_tables}, \ref{fig:SLex3_hypotheses_automata},
and \ref{fig:SLex3_psi}, respectively.

The learner starts asking MQs for the empty word. A symbolic letter $\a_0$ is
chosen to represent its continuations with the minimal element of
$\S$ as evidence, i.e., $\mu(\a_0) = \binom {0} {0}$. The symbolic word $\a_0$ is moved to
$\sy{S}$ for the table $\sy{T}_0$ to be closed.
The symbolic letter $\a_1$ is added to the alphabet of state $\a_0$, and the learner asks a MQ for $\binom {0} {0}
\binom {0} {0}$, the evidence of the symbolic word $\a_0 \a_1$. The first
hypothesis automaton is $\sy{\A}_0$ with $\S$-semantics $[\a_0] = [\a_1] = \S$.
The counter-example $(\binom {45} {50}, -)$ refines the partition
for the initial state. The symbolic alphabet is extended to $\sy{\S}_{\e} =
\{\a_0, \a_2\}$ with $[\a_2] = \{x>=\binom {45}{50}\}$,
$[\a_0] = \S - [\a_2]$, and $\mu(\a_2) = \binom {45}{50}$.
The new observation table and hypothesis are $\sy{T}_1$ and
$\sy{\A}_1$. Two more counter-examples will come to refine the partition for the
initial state, $(\binom {60} {0}, -)$ and $(\binom {0} {70}, -)$, that will
modify the partition for the initial state, moving all letters greater than
$\binom {60} {0}$ and $\binom {0} {70}$ to the $\S$-semantics of
$\a_2$ as can be seen in $\psi_2$ and $\psi_3$ respectively.

After the hypothesis $\sy{\A}_3$, the counter-example ($\binom {0} {0} \binom {0}
{80}, -)$ adds a new symbol $\a_3$ and a new transition
in the hypothesis automaton. The counter-examples that follow, namely, $(\binom
{0} {0} \binom {80} {0},-)$, $(\binom{0}{0}\binom{40}{15},-)$, and
$(\binom{0}{0}\binom{30}{30},-)$ refine the $\S$-semantics for symbols in
$\sy{\S}_{\a_0}$ as shown in $\psi_{4-7}$.

Then counter-example $(\binom {45}{50} \binom {0} {0},+)$ is presented.
As we can see, the prefix $\binom{45}{50}$ exist already in $\mu(\a_2)$ and
$\a_2\in\sy{R}$ which means $\a_2$ becomes a state, and to distinguish it from
the state represented by the empty word the learner adds to $E$ the suffix
of the counter-example $\binom {0}{0}$. The resulting table $\sy{T}_8$ is not closed
and $\a_0\a_1$ is moved to $\sy{S}$. The new table $\sy{T}_9$ is
closed and evidence compatible. The hypothesis $\sy{\A}_9$ has now four states and
the symbolic alphabet and $\S$-semantics for each state can be seen in $\psi_9$.
The counter-examples that follow will refine the partition at state $\a_2$. The
new transitions discovered and all refinements are shown in $\sy{\A}_{10-18}$
and $\psi_{10}-\psi_{18}$. The language was learned using $20$ membership queries
and 17 counter-examples.  \qed
\end{exa}

\section{On Complexity}\label{S:complexity}

The complexity of the symbolic algorithm is influenced not by the size of
the alphabet but by the resolution (partition size) with which we observe it. 
Let $L\subset\S$ be the target language and let $\sy{\A}$ be the minimal 
symbolic automaton recognizing this language with state set $Q$ of size $n$ and a 
symbolic alphabet $\sy{\S} = \biguplus_{q}\sy{\S}_q$ such that $|\sy{\S}_q|\leq m$ 
for every $q$. 

Each counter-example improves the hypothesis in one out of two ways. 
Either a new state is discovered or a partition gets refined. 
Hence, at most $n-1$ equivalence queries of the first type can be asked and $n(m-1)$ of the second, 
resulting in $\mathcal{O}(mn)$ equivalence queries. 

Concerning the size of the table, the set of prefixes $\sy{S}$ is monotonically increasing
and reaches the size of exactly $n$ elements. 
Since the table, by construction, is always kept 
reduced, the elements in $\sy{S}$ represent exactly the states of the
automaton. The size of the boundary is always smaller than the total number of
transitions in the automaton, that is $mn-n+1$. The number of suffixes in $E$, playing a
distinguishing role for the states of the automaton, range between
$\log_2n$ and $n$. Hence, the size of the table ranges between
$(n+m)\log_2n$ and $n(mn + 1)$. 

For a totally ordered alphabet 
the size of the concrete sample coincides with the size of the symbolic sample 
associated with the table and hence the number of membership queries asked is $\mathcal{O}(mn^2)$. 
For a partially ordered alphabet with 
each $F_i$ defined by at most $l$ points, some additional queries 
are asked. For every row in $\sy{S}$, at most $n(m-1)(l-1)$ additional
words are added to the concrete sample, hence more membership queries might
need to be asked. Furthermore, at most $l-1$ more counter-examples are given to
refine a partition. 
To conclude, the number of queries in total asked to learn language $L$
is $\mathcal{O}(mn^2)$ if $l< n$ and $\mathcal{O}(lmn)$ otherwise. 


\section{Conclusion}\label{S:discussion}

We have defined a generic algorithmic scheme for automaton learning, targeting  languages over large alphabets that can be recognized by finite symbolic automata having a modest number of states and transitions. Some ideas similar to ours have been proposed for the particular case of parametric languages  \cite{BergJR06} and recently in a more general setting \cite{alph-refinement,local-alph,bb13sigma} including partial evidential support and alphabet refinement during the learning process.

The genericity of our algorithm is due to a semantic approach (alphabet partitions) but of course, each and every domain will have its own semantic and syntactic specialization in terms of the size and shape of the alphabet partitions. In this work we have implemented an instantiation of this scheme for alphabets such as $(\N, \leq)$ and $(\R,\leq)$. 
When dealing with numbers, the partition into a finite number of intervals (and monotone sets in higher dimensions) is very natural and used in many application domains ranging from quantization of sensor readings to income tax regulations. It will be interesting to compare the expressive power and succinctness of symbolic automata with other approaches for representing numerical time series and to compare our algorithm with other inductive inference techniques for sequences of numbers.

As a first excursion into the domain, we have made quite strong assumptions on the nature of the equivalence oracle, which, already for small alphabets, is a bit too strong and pedagogical to be realistic. We assumed that it provides the shortest counter-example and also that it chooses always the minimal available concrete symbol. We can relax the latter (or both) and even omit this oracle altogether and replace it by random sampling, as already proposed in \cite{Angluin87} for concrete learning.
Over large alphabets, it might be even more appropriate to employ probabilistic convergence criteria a-la \emph{PAC learning} \cite{valiant1984theory} and be content with a correct classification of a large fraction of the words, thus tolerating imprecise tracing of boundaries in the alphabet partitions.
This topic is subject to ongoing work. Another challenging research direction is the adaptation of our framework to languages over Boolean vectors.

\section*{Acknowledgement}
This work was supported by the French project EQINOCS (ANR-11-BS02-004). We thank Peter Habermehl, Eugene Asarin and anonymous referees for useful comments and pointers to the literature.

\bibliographystyle{alpha}
\bibliography{learn-bib}

\newcommand{\etalchar}[1]{$^{#1}$}
\begin{thebibliography}{VHL{\etalchar{+}}12}

\bibitem[Ang87]{Angluin87}
Dana Angluin.
\newblock Learning regular sets from queries and counterexamples.
\newblock {\em Information and Computation}, 75(2):87--106, 1987.

\bibitem[BB13]{bb13sigma}
Matko Botin\v{c}an and Domagoj Babi\'{c}.
\newblock Sigma*: {S}ymbolic learning of {I}nput-{O}utput specifications.
\newblock In {\em POPL}, pages 443--456. ACM, 2013.

\bibitem[BJR06]{BergJR06}
Therese Berg, Bengt Jonsson, and Harald Raffelt.
\newblock Regular inference for state machines with parameters.
\newblock In {\em FASE}, volume 3922 of {\em LNCS}, pages 107--121. Springer,
  2006.

\bibitem[BLP10]{benedikt2010}
Michael Benedikt, Clemens Ley, and Gabriele Puppis.
\newblock What you must remember when processing data words.
\newblock In {\em AMW}, volume 619 of {\em {CEUR} Workshop Proceedings}, 2010.

\bibitem[BR04]{BergR04}
Therese Berg and Harald Raffelt.
\newblock Model checking.
\newblock In {\em Model-Based Testing of Reactive Systems}, volume 3472 of {\em
  LNCS}, pages 557--603. Springer, 2004.

\bibitem[DlH10]{dehiguera2010book}
Colin De~la Higuera.
\newblock {\em Grammatical inference: learning automata and grammars}.
\newblock Cambridge University Press, 2010.

\bibitem[DR95]{Traces}
Volker Diekert and Grzegorz Rozenberg.
\newblock {\em The Book of Traces}.
\newblock World Scientific, 1995.

\bibitem[DV14]{Veanes-minimization}
Loris D'Antoni and Margus Veanes.
\newblock Minimization of symbolic automata.
\newblock In {\em POPL}, pages 541--554. ACM, 2014.

\bibitem[Gol72]{Gold72}
E.~Mark Gold.
\newblock System identification via state characterization.
\newblock {\em Automatica}, 8(5):621--636, 1972.

\bibitem[HJJ{\etalchar{+}}95]{mona95}
Jesper~G. Henriksen, Ole~J.L. Jensen, Michael~E. Jørgensen, Nils Klarlund,
  Robert Paige, Theis Rauhe, and Anders~B. Sandholm.
\newblock Mona: Monadic second-order logic in practice.
\newblock In {\em TACAS}, volume 1019 of {\em LNCS}, pages 80--110. Springer,
  1995.

\bibitem[HSJC12]{HowarSJC12}
Falk Howar, Bernhard Steffen, Bengt Jonsson, and Sofia Cassel.
\newblock Inferring canonical register automata.
\newblock In {\em VMCAI}, volume 7148 of {\em LNCS}, pages 251--266. Springer,
  2012.

\bibitem[HSM11]{alph-refinement}
Falk Howar, Bernhard Steffen, and Maik Merten.
\newblock Automata learning with automated alphabet abstraction refinement.
\newblock In {\em VMCAI}, volume 6538 of {\em LNCS}, pages 263--277. Springer,
  2011.

\bibitem[HV11]{Veanes-vmcai11}
Pieter Hooimeijer and Margus Veanes.
\newblock An evaluation of automata algorithms for string analysis.
\newblock In {\em VMCAI}, volume 6538 of {\em LNCS}, pages 248--262. Springer,
  2011.

\bibitem[IHS13]{local-alph}
Malte Isberner, Falk Howar, and Bernhard Steffen.
\newblock Inferring automata with state-local alphabet abstractions.
\newblock In {\em NASA Formal Methods}, volume 7871 of {\em LNCS}, pages
  124--138. Springer, 2013.

\bibitem[KF94]{KaminskiF94}
Michael Kaminski and Nissim Francez.
\newblock Finite-memory automata.
\newblock {\em Theoretical Computer Science}, 134(2):329--363, 1994.

\bibitem[MM14]{MalerMens14}
Oded Maler and Irini{-}Eleftheria Mens.
\newblock Learning regular languages over large alphabets.
\newblock In {\em TACAS}, volume 8413 of {\em LNCS}, pages 485--499. Springer,
  2014.

\bibitem[Moo56]{moore1956gedanken}
Edward~F Moore.
\newblock Gedanken-experiments on sequential machines.
\newblock In {\em Automata studies}, volume~34 of {\em Annals of Mathematical
  Studies}, pages 129--153. Princeton, 1956.

\bibitem[MP95]{maler1995learnability}
Oded Maler and Amir Pnueli.
\newblock On the learnability of infinitary regular sets.
\newblock {\em Information and Computation}, 118(2):316--326, 1995.

\bibitem[Ner58]{Nerode58}
Anil Nerode.
\newblock Linear automaton transformations.
\newblock {\em Proceedings of the American Mathematical Society},
  9(4):541--544, 1958.

\bibitem[Val84]{valiant1984theory}
Leslie~G. Valiant.
\newblock A theory of the learnable.
\newblock {\em Communications of the ACM}, 27(11):1134--1142, 1984.

\bibitem[VB12]{Veanes-SAtoolkit}
Margus Veanes and Nikolaj Bj{\o}rner.
\newblock Symbolic automata: The toolkit.
\newblock In {\em TACAS}, volume 7214 of {\em LNCS}, pages 472--477. Springer,
  2012.

\bibitem[VHL{\etalchar{+}}12]{Veanes-symb}
Margus Veanes, Pieter Hooimeijer, Benjamin Livshits, David Molnar, and Nikolaj
  Bj\"{o}rner.
\newblock Symbolic finite state transducers: algorithms and applications.
\newblock In {\em POPL}, pages 137--150. {ACM}, 2012.

\end{thebibliography}

\end{document}